\documentclass[twocolumn,preprintnumbers,amsmath,amssymb,showpacs,prl,apsrev4-1]{revtex4-1}
\usepackage{graphicx}
\usepackage{dcolumn}
\usepackage{bm}
\usepackage{SIunits}
\usepackage{verbatim}
\usepackage{placeins}

\begin{document}
\title{Coexistence of localized and itinerant magnetism in newly discovered iron-selenide superconductor LiFeO$_{2}$Fe$_{2}$Se$_{2}$}
\author{Da-Yong Liu$^{1}$, Xiang-Long Yu$^{1,2}$, Ya-Min Quan$^{1}$, Xiao-Jun Zheng$^{1,2}$,
        and Liang-Jian Zou$^{1,3}$ }
\altaffiliation{Corresponding author} \email{zou@theory.issp.ac.cn}
\affiliation{
      $^1$ Key Laboratory of Materials Physics, Institute of Solid State
      Physics, Chinese Academy of Sciences, P. O. Box 1129, Hefei, Anhui
      230031, China \\
      $^2$ University of Chinese Academy of Sciences, Beijing 100000, China\\
      $^3$ Department of Physics, University of Science and Technology of China, Hefei, 230026, China
}

\date{\today}

\begin{abstract}
The electronic structure and magnetism of LiFeO$_{2}$Fe$_{2}$Se$_{2}$ are investigated using the first-principle calculations. The ground state is N$\acute{e}$el antiferromagnetic (AFM) Mott insulating state for Fe1 with localized magnetism in LiFeO$_{2}$ layer and striped AFM metallic state for Fe2 with itinerant magnetism in Fe$_{2}$Se$_{2}$ layer, accompanied with a weak interlayer AFM coupling between Fe1 and Fe2 ions, resulting in a coexistence of localized and itinerant magnetism.
Moreover, the layered LiFeO$_{2}$ is found to be more than an insulating block layer but responsible for enhanced AFM correlation in Fe$_{2}$Se$_{2}$ layer through the interlayer magnetic coupling. The interplay between the magnetisms of Fe1 and Fe2 introduces a control mechanism for spin fluctuations associated with superconductivity in iron-based superconductors.
\end{abstract}


\vskip 300 pt

\maketitle


Since the discovery of superconductivity in layered iron-based superconductor \cite{JACS130-3296}, many iron pnictides and selenides have been found, including 1111 (LaOFeAs and SmOFeAs, $\it{etc.}$), 111 (LiFeAs and NaFeAs, $\it{etc.}$), 122 (BaFe$_{2}$As$_{2}$ and KFe$_{2}$Se$_{2}$, $\it{etc.}$) and 11 (FeSe and FeTe) systems \cite{Nat453-761,PRB82-180520,RMP85-849,PRB78-134514,nphy8-709,PRL102-247001}.
Especially, FeSe and its intercalated compounds A$_{x}$Fe$_{2-y}$Se$_{2}$ (A= alkali metal, K, Rb and Cs, $\it{etc.}$) have been extensively investigated\cite{PRL102-177003,PRL106-187001,PRL107-056401,JPCM25-125601}.
Due to the proximity of antiferromagnetism and superconductivity in these iron-based compounds, the magnetism has become a central issue. One of the hotly debated topics, the origin of magnetism, ${\it i.e.}$, whether it comes from Fermi surface (FS) nesting mechanism of itinerant picture or frustrated superexchange one of localized picture, has attracted great interest \cite{EPL83-27006,PRL100-237003,PRL105-107004,nmat10-932,RMP85-849}. Therefore, the coexistence of localized and itinerant magnetism and superconductivity observed in iron-based compounds provides an opportunity to better understand the roles of the localized and itinerant electrons \cite{nphy8-709,PRB84-054527}, such as in the SmFeAsO$_{1-x}$F$_{x}$ and CeFeAsO$_{1-x}$F$_{x}$ with high superconducting transition temperature T$_{c}$ \cite{Nat453-761,nmat8-310,PRB84-054419,nmat7-953}.

Recently, a novel iron selenide compound LiFeO$_{2}$Fe$_{2}$Se$_{2}$ with high superconducting transition temperature T$_{c}$ up to 43 K is synthesised \cite{PRB89-020507}. It is isostructural with ROFeAs (R = rare earth, La, Sm, $\it{etc.}$) compound. Experimentally, the anti-PbO-type layers of LiFeO$_{2}$ have been intercalated between anti-PbO-type Fe$_{2}$Se$_{2}$ layers. This leads to a very large crystal lattice parameter $c$, making the Fe$_{2}$Se$_{2}$ layer two-dimensional. There are two kinds of Fe ions in different layers, $\it{i.e.}$ Fe1 ions in LiFeO$_{2}$ layer and Fe2 ions in Fe$_{2}$Se$_{2}$ layer, very different from the other iron-based superconductors. Experimentally, bulk LiFeO$_{2}$ compound is an antiferromagnetic (AFM) insulator with large magnetic moment \cite{JSSC140-159}, suggesting a possible coexistence of localized and itinerant magnetisms in the newly discovered compound LiFeO$_{2}$Fe$_{2}$Se$_{2}$.

In this Letter, to clarify the roles of the two different kinds of Fe ions in LiFeO$_{2}$Fe$_{2}$Se$_{2}$, we performed the first-principle calculations and disentangled the 3$d$-bands of Fe2 from the other bands using Wannier functions. We find that in the nonmagnetic phase, both Fe1 ions in LiFeO$_{2}$ layer and Fe2 ions in Fe$_{2}$Se$_{2}$ layer contribute to the low energy band structures and the unique FS topology. The magnetic ground state for Fe2 ions in Fe$_{2}$Se$_{2}$ layer is striped AFM (SAFM) state displaying a bad metallic behavior, like that of LaOFeAs; while for Fe1 ions in LiFeO$_{2}$ layer, it is a N$\acute{e}$el AFM (NAFM) state displaying a Mott insulating behavior. These results show a novel scenario obviously different from other iron-based superconductors, which may provide a new reference material to clarify the roles of nesting and magnetic frustration.
%

%
In our calculations, the full potential linearized augmented-plane-wave (FP-LAPW) scheme based on density functional theory (DFT) in the code WIEN2K package \cite{WIEN2K} was used. Exchange and correlation effects are taken into account in the generalized gradient approximation (GGA) by Perdew, Burk, and Ernzerhof (PBE) \cite{PBE}. In order to calculate the magnetic structure, the 1 $\times$ 1 $\times$ 1 unit cell, $\sqrt{2} \times \sqrt{2} \times 1$ supercell, and $2 \times 2 \times 1$ supercell are used for nonmagnetic (NM) and N$\acute{e}$el AFM (NAFM) sates, striped AFM (SAFM) state, and bi-collinear AFM (BAFM) state, respectively. A sufficient number of k points is used, 26 $\times$ 26 $\times$ 11 for nonmagnetic calculations, and 11 $\times$ 11 $\times$ 6 for magnetic supercell calculations. In the disentanglement procedure, the maximally localized Wannier functions (MLWF) scheme, implemented with WANNIER90 \cite{CPC178-685} and WIEN2WANNIER \cite{CPC181-1888}, is used.

\begin{figure}[htbp]
\hspace*{-2mm}
\centering
\includegraphics[trim = 0mm 0mm 0mm 0mm, clip=true, width=1.0 \columnwidth]{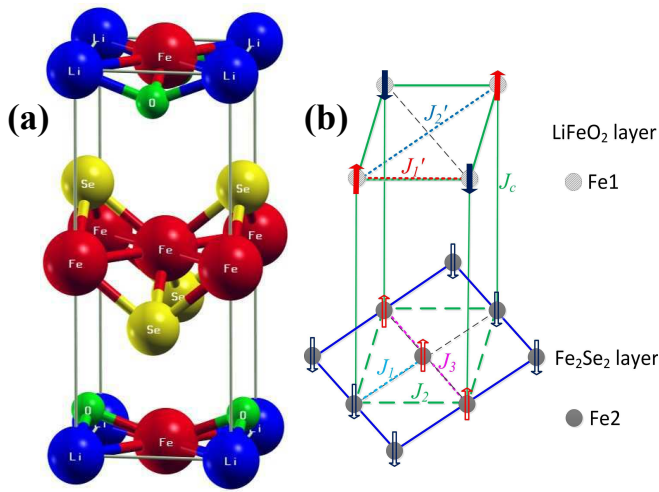}
\caption{(Color online) Crystal structure (a) with Li, Fe, O and Se atoms in blue, red, green, and yellow, respectively, and schematic representations of the magnetic structure (b) of LiFeO$_{2}$Fe$_{2}$Se$_{2}$.}
\label{str}
\end{figure}

%
In order to compare our numerical results with the experimental data, we adopt the experimental structural data of LiFeO$_{2}$Fe$_{2}$Se$_{2}$ given by X-ray diffraction \cite{PRB89-020507}. LiFeO$_{2}$Fe$_{2}$Se$_{2}$ has tetragonal structure with space group $P$-$4m2$ (No. 115) and lattice parameters $a$ = 3.7926 ${\AA}$, and $c$ = 9.2845 ${\AA}$ \cite{PRB89-020507}, as shown in Fig.~\ref{str}. In Fe$_{2}$Se$_{2}$ layer, the Fe ions form a square lattice as in the other iron-based compounds. The nearest-neighbor (N.N.) Fe2-Fe2 distance is about 2.6818 $\AA$, and the next nearest-neighbor (N.N.N.) distance is about $\sqrt{2}$ times of the N.N. distance, 3.7926 $\AA$. And the Fe2 is surrounded by four Se atoms which form a tetrahedron crystal field environment. While in LiFeO$_{2}$ layer, Fe1 ions also form a $\sqrt{2}$ $\times$ $\sqrt{2}$ square lattice compared with that in Fe$_{2}$Se$_{2}$ layer.

\begin{figure}[htbp]\centering
\includegraphics[angle=0, width=1.0 \columnwidth]{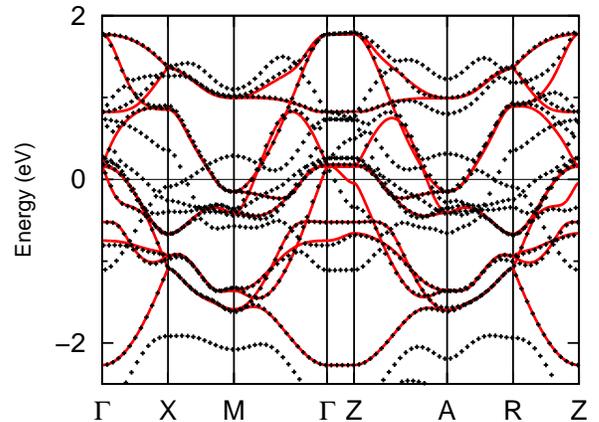}
\caption{(Color online) Band structures of nonmagnetic state in LiFeO$_{2}$Fe$_{2}$Se$_{2}$. The plus symbols represents the band structure obtained by GGA, and the solid lines show the band structures projected on the 3d orbitals of Fe2 using MLWF.} \label{wfband}
\end{figure}

\begin{figure}[htbp]\centering
\includegraphics[angle=0, width=1.0 \columnwidth]{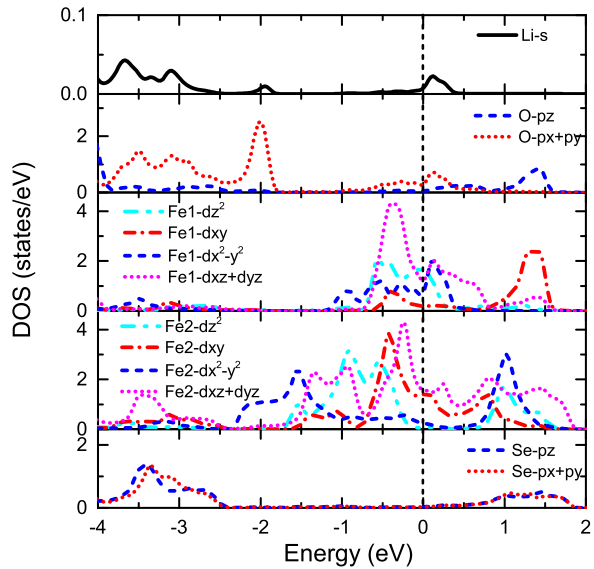}
\caption{(Color online) Partial density of states of nonmagnetic LiFeO$_{2}$Fe$_{2}$Se$_{2}$.} \label{npdos}
\end{figure}

We first explore the nonmagnetic phase of LiFeO$_{2}$Fe$_{2}$Se$_{2}$, which can provide a comparison for the magnetic case. The band structures of nonmagnetic state in LiFeO$_{2}$Fe$_{2}$Se$_{2}$ within GGA are shown in Fig.~\ref{wfband}.
The band structure of 3$d$ orbitals for Fe2 is disentangled from the other bands using MLWF. From the band structure in Fig.~\ref{wfband} and the partial density of states (PDOS) in Fig.~\ref{npdos}, it is obviously found that the 3$d$-bands of both Fe1 and Fe2 cross the Fermi level.
%

\begin{figure}[htbp]\centering
\includegraphics[trim = 0mm 0mm 0mm 0mm, clip=true, width=1.0 \columnwidth]{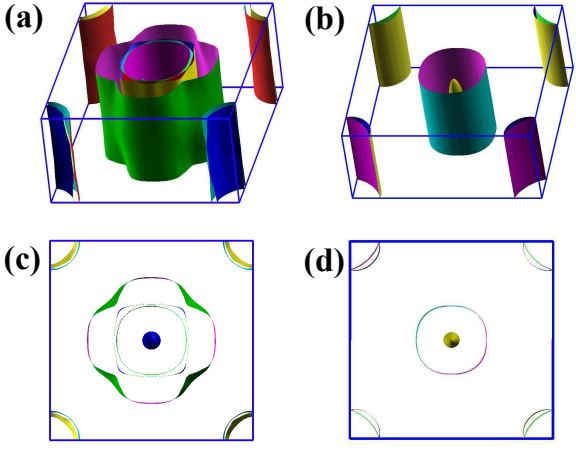}
\caption{(Color online) Fermi surface (a) and its top view (c) obtained by GGA, and Fermi surface projected on Fe2 (b) and its top view (d) obtained by MLWF in nonmagnetic LiFeO$_{2}$Fe$_{2}$Se$_{2}$.
}\label{nfs}
\end{figure}

The unique FS structure and topology in LiFeO$_{2}$Fe$_{2}$Se$_{2}$, as shown in Fig.~\ref{nfs}, are very different from other iron-based superconductors \cite{PRB84-064435,physicab407-1139,JPCM25-125601}. It mainly consists of a small hole pockets contributed from Fe2 at $\Gamma$ point, two inner hole cylinders contributed from Fe2, two outer hole cylinders contributed from Fe1, and two electron cylinders contributed from Fe2 at M or A points. Due to the large $c$ lattice parameter, the Fe$_{2}$Se$_{2}$ layer is nearly separated from the LiFeO$_{2}$ layer. Thus the FS can be seen as a composite structure, which is contributed from Fe1 and Fe2, respectively.

\begin{figure}[htbp]\centering
\includegraphics[trim = 0mm 0mm 0mm 0mm, clip=true, width=1.0 \columnwidth]{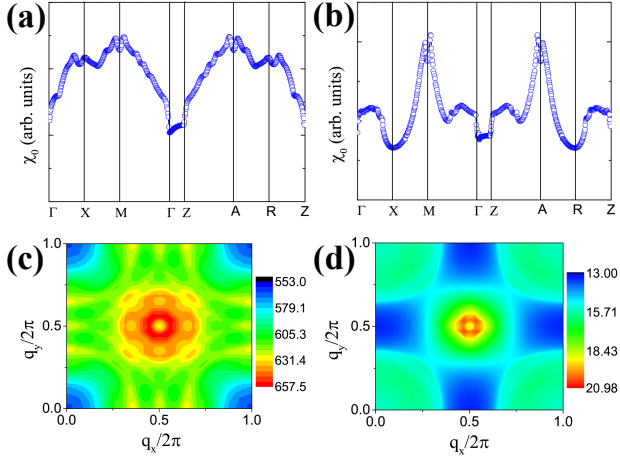}
\caption{(Color online) Dynamical susceptibilities along high symmetry path and q$_{x}$-q$_{y}$ plane with q$_{z}$ = 0 (left up and down panels) of LiFeO$_{2}$Fe$_{2}$Se$_{2}$, and the contributions from only 3$d$ orbitals of Fe2 (right up and down panels) in Fe$_{2}$Se$_{2}$ layer are also given for comparison.
}\label{chi}
\end{figure}

In order to clearly explore whether there is FS nesting in LiFeO$_{2}$Fe$_{2}$Se$_{2}$, we study the dynamical spin susceptibility of LiFeO$_{2}$Fe$_{2}$Se$_{2}$. The spin susceptibility is given by \cite{NJP11-025016,physicab407-1139}
\begin{eqnarray}
  \chi_{0}(q)=\frac{1}{N}\sum_{\substack{\vec{k},n,m}}
  \frac{f(\varepsilon_{n}(\vec{k}))-f(\varepsilon_{m}(\vec{k}+\vec{q}))}
  {\varepsilon_{m}(\vec{k}+\vec{q})-\varepsilon_{m}(\vec{k})+i\eta}
\end{eqnarray}
The obtained dynamical susceptibilities are plotted in Fig.~\ref{chi}.
It clearly shows two relatively broadened peaks, one appears at near M point with $Q \sim (\pi,\pi,\pi)$, and another is relatively small at near A point with $Q \sim (\pi,\pi,0)$.
There is no very sharp peak, indicating the FS nesting is poor in LiFeO$_{2}$Fe$_{2}$Se$_{2}$.
For comparison, the spin susceptibility contributed from only 3$d$ orbitals of Fe2 in Fe$_{2}$Se$_{2}$ layer is also given, which displays a relatively sharp peak at M and A points, suggesting the perfect nesting is determined by the 3$d$ orbitals of Fe2 in Fe$_{2}$Se$_{2}$ layer, similar to the other iron-based superconductors \cite{physicab407-1139}. Therefore the imperfect nesting in LiFeO$_{2}$Fe$_{2}$Se$_{2}$ may arise from the $p$-$d$ hybridizations between Fe2 3$d$ and Se 4$p$ orbitals, and the interlayer couplings between LiFeO$_{2}$ and Fe$_{2}$Se$_{2}$ layers.

\begin{figure}[htbp]\centering
\includegraphics[trim = 0mm 0mm 0mm 0mm, clip=true, width=1.0 \columnwidth]{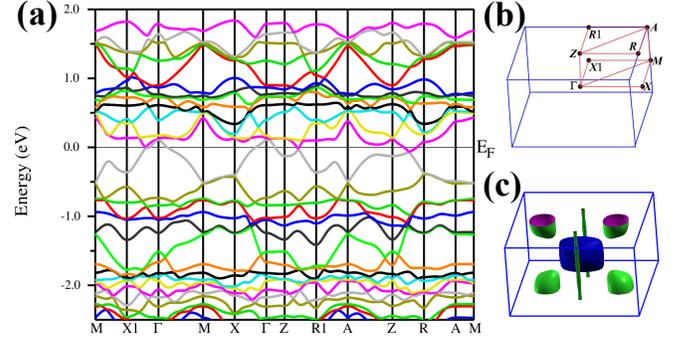}
\caption{(Color online) The band structures (a), Brillouin zone (b) and Fermi surface (c) of the magnetic ground state in LiFeO$_{2}$Fe$_{2}$Se$_{2}$ obtained by GGA.
}\label{mbandfs}
\end{figure}

%
To search for the magnetic ground state of LiFeO$_{2}$Fe$_{2}$Se$_{2}$, several magnetic structures are investigated, including NM, ferromagnetic (FM), NAFM, SAFM and BAFM states in Fe$_{2}$Se$_{2}$ layer, NM, FM and NAFM states in LiFeO$_{2}$ layer, and with interlayer FM or AFM couplings. We found that the magnetic ground state is SAFM in Fe$_{2}$Se$_{2}$ layer and NAFM in LiFeO$_{2}$ layer, accompanied with interlayer AFM coupling, as shown in Fig.~\ref{str}, in accordance with the nesting vector $Q = (\pi,\pi,\pi)$. Fig.~\ref{mbandfs} shows the band structure and FS in the magnetic ground state. The system becomes a bad metal in comparison with the nonmagnetic phase. We notice that due to the band-crossing points locating nearly at Fermi level along $\Gamma$-$X1$ and $Z$-$R1$ paths in the magnetic Brillouin zone, the Fermi surface can be easily affected by doping, electronic correlation or lattice distortion.

\begin{figure}[htbp]\centering
\includegraphics[angle=0, width=1.0 \columnwidth]{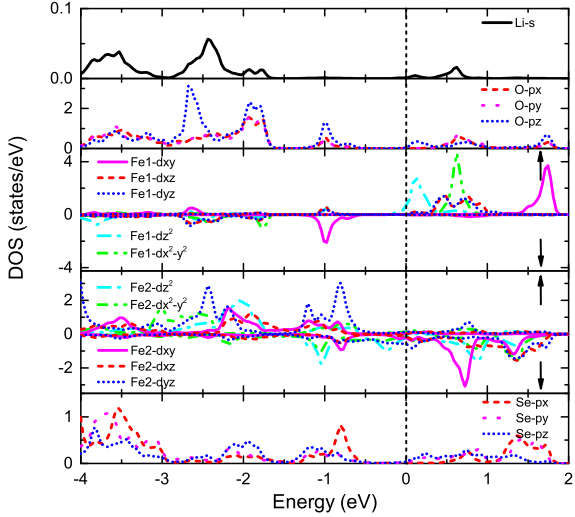}
\caption{(Color online) Partial density of states of magnetic ground state in LiFeO$_{2}$Fe$_{2}$Se$_{2}$.} \label{mpdos}
\end{figure}

The PDOS in the magnetic ground state is plotted in Fig.~\ref{mpdos}. It is clear that the Fe1$^{3+}$ (3$d^{5}$) ions give a nearly semiconducting state with localized magnetism. Further considering the Coulomb electronic correlation, the states would become fully gaped within GGA+$U$ framework. Notice that for bulk LiFeO$_{2}$ compound, both the experiment \cite{JSSC140-159} and GGA+$U$ calculation \cite{JMMM343-92} suggest a Mott insulator with N$\acute{e}$el AFM.
Meanwhile the Fe2$^{2+}$ (3$d^{6}$) ions display a bad metallic behavior with itinerant magnetism, resulting in a coexistence of localized and itinerant magnetisms, similar to SmFeAsO$_{1-x}$F$_{x}$ \cite{Nat453-761}.
Moreover, because of the large $c$ lattice parameter, there is nearly no charge transfer between the LiFeO$_{2}$ and Fe$_{2}$Se$_{2}$ layers. Therefore, the LiFeO$_{2}$ layer is an insulating block one, while the Fe$_{2}$Se$_{2}$ layer is a conducting one.

In order to describe the coexistence of the localized and itinerant magnetisms in this system, a Heisenberg model is constructed as follows,
\begin{equation}
\label{eq.1}
\begin{aligned}
H=&J_{1}\sum_{\substack{<ij>_{Fe2}}}\vec{s}_{i}\cdot\vec{s}_{j}
+J_{2}\sum_{\substack{<<ij>>_{Fe2}}}\vec{s}_{i}\cdot\vec{s}_{j}\\
\nonumber
&+J_{3}\sum_{\substack{<<<ij>>>_{Fe2}}}\vec{s}_{i}\cdot\vec{s}_{j}
+J^{'}_{1}\sum_{\substack{<mn>_{Fe1}}}\vec{S}_{m}\cdot\vec{S}_{n}\\
\nonumber
&+J^{'}_{2}\sum_{\substack{<<mn>>_{Fe1}}}\vec{S}_{m}\cdot\vec{S}_{n}
+J_{c}\sum_{\substack{<im>_{c}}}\vec{S}_{m}\cdot\vec{s}_{i},
\end{aligned}
\end{equation}
where $J_{1}$, $J_{2}$ and $J_{3}$ are the N.N., N.N.N. and third-nearest-neighbor intralayer magnetic couplings in Fe$_{2}$Se$_{2}$ layer with spin $s$ of Fe2 ions, $J^{'}_{1}$ and $J^{'}_{2}$ the N.N. and N.N.N. intralayer magnetic couplings in LiFeO$_{2}$ layer with spin $S$ of Fe1 ions, and $J_{c}$ the N.N. interlayer magnetic coupling between Fe1 and Fe2 ions. The spin exchange parameters can be obtained by differences of total energies of different magnetic structures, which are listed in the following,
\begin{equation}
\label{eq.2}
\begin{aligned}
\Delta E(NAF-BAF)_{Fe_{2}}&=&&-4(J_{1}-J_{2}-2J_{3})s^{2} \\
\nonumber
\Delta E(NAF-SAF)_{Fe_{2}}&=&&-4(J_{1}-2J_{2})s^{2} \\
\nonumber
\Delta E(NAF-FM)_{Fe_{2}}&=&&-8J_{1}s^{2} \\
\nonumber
\Delta E(NAF-FM)_{Fe_{1}}&=&&-8J^{'}_{1}S^{2}-2J_{c}Ss \\
\nonumber
\Delta E(NAF-SAF)_{Fe_{1}}&=&&-4(J^{'}_{1}-2J^{'}_{2})S^{2}-2J_{c}Ss \\
\nonumber
\Delta E(AF-FM)_{Fe_{1,2}}&=&&-4J_{c}Ss
\end{aligned}
\end{equation}
The calculated spin exchange parameters are $J_{1}$ = 27.67 meV/$s^{2}$,
$J_{2}$ = 19.74 meV/$s^{2}$, $J_{3}$ = 3.35 meV/$s^{2}$, $J^{'}_{1}$ = 49.61 meV/$S^{2}$,
$J^{'}_{2}$ = 6.24 meV/$S^{2}$, and $J_{c}$ = 2.29 meV/$Ss$. Considering the calculated magnetic moments 3.5 $\mu_{B}$ for Fe1 and 2.4 $\mu_{B}$ for Fe2, we adopt Fe1 spin S=2 and Fe2 spin s=3/2. The positive values of all the spin exchange parameters favor AFM coupling. Moreover, the relationship $J_{2}>J_{1}/2$ results in a strong magnetic frustration in Fe$_{2}$Se$_{2}$ layer of LiFeO$_{2}$Fe$_{2}$Se$_{2}$, similar as in LaOFeAs. Consequently the magnetic structure can be well understood within this Heisenberg model. In addition, we have also compared the magnetic case of LiFeO$_{2}$ with the nonmagnetic one, and found that the strong AFM interaction of LiFeO$_{2}$ layer enhances the AFM correlation in Fe$_{2}$Se$_{2}$ layer through the interlayer magnetic coupling. Thus the strong magnetic fluctuation induced by LiFeO$_{2}$ layer is possibly responsible for the high superconducting transition temperature T$_{c}$ in LiFeO$_{2}$Fe$_{2}$Se$_{2}$, except for the strong two-dimensional character. In fact, the enhancement of magnetism is also observed experimentally in the other iron-based materials, which is ascribed to the influence of the interstitial Fe \cite{PRL107-216403,PRL108-107002}. Therefore our results demonstrate that the interplay between the magnetisms of different layers is a new control mechanism for spin fluctuations associated with superconductivity in the iron-based superconductors.

It is worth noting that there are many sensitive factors which may seriously affect the magnetism in LiFeO$_{2}$Fe$_{2}$Se$_{2}$. One is the possible stoichiometric problem extensively existed in iron-based superconductors, $\it{e.g.}$ excess Fe in the experiments leads to contradictory results between theory and experiment, $\it{i.e.}$ no magnetism was observed experimentally for FeSe \cite{PRL102-177003,nmat8-630} and LiFeAs \cite{PRB78-094511,PRB81-140511}. Another is the cation-disorder of Li and Fe ions in bulk $\alpha$-LiFeO$_{2}$ \cite{JSSC140-159}. Although $\alpha$-LiFeO$_{2}$ remains strong AFM observed experimentally \cite{JSSC140-159}, we can not exclude the possibility that the magnetic ordering may be destroyed by the disorder in LiFeO$_{2}$Fe$_{2}$Se$_{2}$. Thus further neutron scattering and nuclear magnetic resonance (NMR) experiments are deserved to investigate the magnetism in LiFeO$_{2}$Fe$_{2}$Se$_{2}$.

%
In summary, we have performed first-principle calculations for the electronic structure and magnetic properties of LiFeO$_{2}$Fe$_{2}$Se$_{2}$. We find that the low energy physics of the novel iron selenide superconductor LiFeO$_{2}$Fe$_{2}$Se$_{2}$ are dominated by the Fe$_{2}$Se$_{2}$ layer, while the layered LiFeO$_{2}$ layer is not only a Mott insulating block layer, but also responsible for the enhancement of the antiferromagnetic correlations in Fe$_{2}$Se$_{2}$ layer. The coexistence of the localized and itinerant magnetisms in LiFeO$_{2}$Fe$_{2}$Se$_{2}$ provide a good example for the investigation of the interplay of localized and itinerant electrons, or the interplay of the magnetism and superconductivity.

%
We acknowledge the Prof. X.-H. Chen provide us the experimental results.
This work was supported by the National Sciences Foundation of China
under Grant Nos. 11074257, 11104274, 11274310 and 11204311, and Knowledge Innovation Program of the Chinese Academy of Sciences. Numerical calculations were
performed at the Center for Computational Science of CASHIPS. The crystal structure and Fermi surface were visualised with XCrysDen \cite{xcrys}.

\end{document}